# La importancia de realizar estrategias de marketing en el ecosistema de emprendimiento universitario

## *The importance of carrying out marketing strategies in the university entrepreneurship ecosystem*

## *A importância da realização de estratégias de marketing no ecossistema de empreendedorismo universitário*

**Guillermo José Navarro Del Toro**
Universidad de Guadalajara, Centro Universitario de los Altos, México
guillermo.ndeltoro@academicos.udg.mx
https://orcid.org/0000-0002-4316-879X

**Resumen**

Sin saber la fecha de retorno a actividades presenciales fue un reto para implementar las estrategias que deberían de efectuarse para que, con bases en las patentes en que han estado trabajando algunos investigadores del CUAltos a la cuales, se les han estado incorporando estrategias de marketing para lograr incrementar su alcance de mercado, fue la ocasión más propicia para que fueran tomadas como base para impulsar al estudiante y empresario a que incursionen en emprendimientos propios que al contar con la asesoría por parte del personal del CIIO, tuvieran más probabilidades de ser exitosos.

La pandemia de SRAS-COV2, marcó un periodo caracterizado por la reducción de todas las actividades que realiza ser humano. En la educación de nivel superior, no se tenía idea de la fecha para el retorno frente a grupo, por lo que se dejaron de hacer una gran cantidad de actividades propias de las universidades. En este proyecto se hace énfasis en las actividades entorno del emprendimiento, ya que fueron casi nulas, por ello, al regresar frente a grupo se realizaron muchas actividades para fomentar e impulsar el emprendimiento, en donde participa el CIIO del CUAltos de tratando de recuperar el tiempo en que no se pudo hacer emprendimiento, por lo que se implementaron programas para fomentar, como la obtención de la certificación ISO 9001:2015 , para incrementar los proyectos de emprendedores que sean patentables y contribuyen al desarrollo de la Región de los Altos de Jalisco y del propio CUAltos.

**Palabras clave:** asesoría, emprendimiento, IES, patente, marketing.

## Abstract

Without knowing the date of return to face-to-face activities, it was a challenge to implement the strategies that should be carried out so that, based on the patents that some researchers from CU Altos have been working on, to which marketing strategies have been incorporated. In order to increase their market reach, it was the most propitious occasion for them to be taken as a basis to encourage students and entrepreneurs to venture into their own ventures that, having the advice of CIIO staff, would be more likely to be successful.

The SARS-COV2 pandemic marked a period characterized by the reduction of all the activities carried out by human beings. In Higher Level Education, there was no idea of the date for the return to the group, so a large number of university activities were stopped. In this project, emphasis is placed on the activities surrounding the entrepreneurship, since they were almost nil, therefore, when returning in front of the group, many activities were carried out to encourage and promote entrepreneurship, in which the CIIO of the CUAltos participates in trying to recover the time when entrepreneurship could not be carried out, so programs were implemented to promote,such as obtaining the ISO 9001:2015 certification on entrepreneurship, to increase entrepreneurship projects that are patentable and contribute to the development of the Los Altos de Jalisco Region and CUAltos itself.

## Resumo

Sem saber a data de regresso às atividades presenciais, foi um desafio implementar as estratégias que deveriam ser levadas a cabo para que, com base nas patentes que alguns investigadores da CUAltos têm vindo a trabalhar, para as quais, têm vindo a ser desenvolvidas estratégias de marketing incorporados para ampliar seu alcance de mercado, foi a ocasião mais propícia para que fossem tomados como base para incentivar estudantes e empreendedores a se aventurarem em seus próprios empreendimentos que, tendo a assessoria da equipe do CIIO, teriam maiores chances de sucesso.

A pandemia de SARS-COV2 marcou um período caracterizado pela redução de todas as atividades realizadas pelo ser humano. No ensino superior, não havia ideia da data para o retorno diante do grupo, então um grande número de atividades típicas das universidades foram paralisadas. Neste projeto é dado destaque às atividades envolventes ao empreendedorismo, uma vez que foram quase nulas, pelo que, ao regressar à frente do grupo, foram realizadas muitas atividades de fomento e fomento ao empreendedorismo, onde o CIIO da CUAltos participa na tentativa de recuperar o tempo em que o empreendedorismo não podia ser realizado, por isso foram implementados programas para promover, como a obtenção da certificação ISO 9001:2015, para aumentar os projetos de empreendedores que são patenteáveis e contribuem para o desenvolvimento de Los Altos de Jalisco Região e da própria CUAltos.

**Palavras-chave:** consultoria, empreendedorismo, IES, patente, marketing.

---

# Introducción

Actualmente, es muy frecuente escuchar que orientar la educación hacia la adquisición de conocimientos ya no es importante, lo cual reside en el hecho de que todo está al alcance en Internet, así lo han expresado Terrazas y Silva (2013). Hay quienes defienden esta postura y añaden que lo importante realmente es aprender a resolver problemas, desarrollar el pensamiento crítico y fomentar la creatividad. Sin embargo, hay que remontarse a lo que la ciencia ha averiguado sobre cómo se desarrollan este tipo de habilidades tan deseables: el desarrollo de la creatividad, la resolución de problemas y el análisis crítico, entre otras habilidades, dependen fundamentalmente de la adquisición de conocimientos, mismos que han sido expresados por Mackay, Franco y Villacis (2018).

Por ello, es que se pretende que estos conocimientos sean significativos de acuerdo con Alvarado, García y Castellanos (2017), es decir, estar bien conectados entre ellos y organizados alrededor de grandes ideas, lo que implica que estén dotados de comprensión y sean transferibles a nuevas situaciones. En otras palabras, deben ser conocimientos profundos. Ya que son los que marcan la diferencia entre la persona experta y el principiante, sin importar el área de conocimientos de que se trate. Lo cual conduce a realizar un análisis de lo que se cree que es la investigación y la forma en que influye en la educación universitaria y los resultados que se pueden desprender de ella.

De acuerdo con Vázquez Parra (2021), existe la creencia de que la investigación universitaria va de la manos con fuertes inversiones monetarias, ya que se hace presente en la creación de grandes laboratorios donde participan investigadores. Sin embargo, contar con un presupuesto de respaldo, es muy conveniente, pero si se carece ideas originales o no se plantean problemas reales que se puedan comprender mejor a través de la investigación, se tendrán inversiones improductivas.

Por ello, Rivadeneira y Silva (2017) plantean el hecho de que es que en muy frecuente ver que el punto clave de las universidades que generan investigación está relacionada con sus estudiantes, quienes, al contar con la debida orientación, pueden ser la fuente que detone el conocimiento original, que se sustente en reflexiones de alto nivel y, además sean de gran valor académico.

El aprendizaje basado en la investigación se convierte en una técnica didáctica cuyo objetivo es relacionar las enseñanzas obtenidas en el aula, con técnicas y metodologías de investigación que permitan que el estudiante pueda, simultáneamente con su formación académica, desarrollar competencias y habilidades de análisis, reflexión y argumentación, con ello, podrá ser partícipe en la construcción del conocimiento lo que concuerda con lo expresado en Tecnológico de Monterrey (2020).

Sin embargo, no solamente por tener más conocimientos, sino por el hecho de que gracias a ellos se puede percibir, interpretar, organizar y emplear la información que recibe de una forma muy distinta a como lo hacen quienes no los tienen tal como puede ser encontrado en Poblete et al. (2019). Esto se traduce en una enorme ventaja para aprender, razonar, crear soluciones y resolver problemas en su disciplina, así como en una mayor capacidad para transferir sus conocimientos y habilidades a otras áreas de conocimiento o de su vida personal y profesional.

Se debe tomar en consideración que el investigador, como una persona experta (en algún área del conocimiento), generalmente, puede detectar patrones que el principiante (el común de la gente) no puede percibir. Lo se debe a que el experto ha integrado conjuntos de datos e ideas en unidades mayores que para él, tienen significado que lo conduce al desarrollo de métodos y productos innovadores que propician el crecimiento de la institución o empresa en donde esté integrado.

Se debe hacer hincapié que la presencia de esos expertos es imprescindible en la universidad y empresa, ya que la economía actual está basada en lo que puede llamarse la *exigencia de innovación,* como es expresado por Ruiz de Olano & Ageitos (2013), lo cual es muy simple de entender, ya que si la empresa (o universidad) no cambia lo que ofrece al mercado (productos y servicios) y no los crea, corre el riesgo de ser superada por otras que sí lo hacen, y puede llegar a desaparecer. En este sentido, la capacidad de innovar se ha convertido en algo cada vez más importante para la creación y mantenimiento de la ventaja competitiva, lo que reside en el hecho de que la compañía (llámese universidad o empresa) entre más innovadora sea, tiende a ser más rentable, con mayor valor de mercado, mejores calificaciones crediticias y probabilidad de supervivencia más alta.

Por ello, universidad y empresa se enfrenta al reto de generar esa innovación, lo que obliga a contar con un departamento que esté orientado a la *investigación*, *desarrollo* e *innovación*, pensando que invirtiendo dinero y recursos tendrá mayor probabilidad de lanzar nuevos y mejores productos al mercado. Para la universidad, implica que contar con patentes, carreras y actividades novedosas es la forma de atraer más investigadores, estudiantes e inversiones que permitan su expansión hacia fuera de la misma, mientras que en la empresa, significa un crecimiento y estar en el gusto y preferencia del cliente. La innovación puede venir desde dentro de la empresa, o bien provenir de la innovación abierta (open innovation) que proviene de la misma universidad. Este tipo de innovación surge en 2003 y de acuerdo con Chesbrough, Vanhaverbeke y West (2014) es un nuevo paradigma que propone que la innovación debe abrirse al exterior de la empresa, desde la apertura de sus procesos de generación de conocimiento hasta la forma de vender sus productos y servicios, según sea el caso.

Es decir que, para hacerlo llegar a una mayor cantidad potencial de público consumidor, se debe tener un sistema muy robusto de marketing, lo que es de suma relevancia para hacer llegar el producto (bien o servicio) a una mayor mercado potencial de consumidores en el nicho de mercado al que esté orientado ese producto, como lo menciona Tovar Góngora (2014). Por lo tanto, es muy recomendable que el sistema de marketing que se use con tal propósito, está basado en metodologías que estén basadas en las tecnologías de la información y comunicación (TIC's) para llegar a una mayor cantidad de consumidores potenciales. Ello implica que, se puede desarrollar el mejor producto o servicio del momento, pero sin una estrategia para hacerlo atractivo al consumidor final, éste será solamente un producto de lo tantos que existen en el mercado.

Pero cuando la empresa carece de un sistema de marketing basado, de ser posible en las últimas versiones del mismo, entonces, estará disminuida su capacidad de dar a conocer su producto o servicio al público consumidor que se encuentre más allá de su área de influencia, lo que influye en una reducción de posibles ganancias y crecimiento. Por su parte, en la universidad cuando no se cuenta con un sistema de marketing que haga posible que se den a conocer sus logros, programas de estudios de carreras de futuro prometedor, emprendimientos que realicen estudiantes, docentes e investigadores, patentes, logros en concursos académicos, divulgación científica, publicaciones en revistas y congresos, asesorías a empresas y empresarios, estudiantes que representen a la universidad, asesoría al emprendedor, entre otros, entonces, será un desperdicio de esfuerzos que se realicen ya que sus productos tendrán un alcance muy limitado.

## Marco Teórico

La investigación, de acuerdo autores como Catalán Cueto (2020), puede ser definida como una actividad humana muy trascendental para la sociedad. No se puede proceder a realizarla sin contar con los antecedentes que conduzcan a tratar de encontrar resultados que brinden resultados a la sociedad en general. Por ello, es necesario conocer hechos, causas relaciones y consecuencias en toda fase del proceso, debiendo hacerse en plena conciencia de todos sus elementos y factores para lograr la eficacia. La investigación debe despertar curiosidad, reflexión, cuestionamiento y duda. De allí la investigación permite que los participantes involucrados desarrollen nuevas formas de comprensión para emprender caminos de reflexión autónoma y compartida sobre el sentido de la práctica y las posibilidades de mejorarla.

En el contexto educativo puede considerarse es una actividad ética que requiere de reflexión y cuestionamiento continuos, por lo que no puede reducirse a una actividad técnica, por ser un proceso que tiene mucha profundidad del proceso, ya que en ella participan docente-estudiante-comunidad, el docente actúa con todo su comportamiento que está enmarcado en sus creencias, actitudes, costumbres y entorno.

Por ello, es muy importante que un investigador, en lo posible, sea simultáneamente docente, para contar con estudiantes de características únicas, que les permitirán aprender algunos conocimientos del investigador, tal como lo ha expresado Diego (2019). Además, es bien sabido que el investigador como lo expresan autores como Sciarelli, Gheith y Tani (2020), generalmente, es un generador de ideas que son base para sus investigaciones, y algunas de ellas no sólo se convertirán en artículos que son publicados en revistas o libros, sino por el contrario, que pueden convertirse en productos con calidad de patentables y por ende, además de reconocimiento a él, universidad y los que hayan participado en la elaboración de ese patentable, podrán estar los de tipo monetario.

Por su parte el estudiante, recibe información y la procesa de acuerdo con sus experiencias, costumbres y entorno, por ello el aprendizaje es una situación incierta, única, cambiante y compleja. El docente, interviene en un escenario cambiante definido por la interacción simultanea de múltiples factores (sociales, económicos, culturales, políticos, entre otros) y condiciones.

De ahí que, la *investigación experimental* de acuerdo con la OCDE (2015), al iniciar con una idea de emprendimiento, tendrá más probabilidades de éxito dependiendo de la habilidad que se tenga al manejar la complejidad y resolver problemas prácticos o situaciones problemáticas. Para ello, se debe recurrir al proceso de reflexión en la acción o una conversación reflexiva con la situación problemática concreta que permitirá crear nuevas realidades, corregir e inventar, o simplemente actuar de forma inteligente y creativa o poner en acción las ideas pertinentes después de su reflexión ante los conflictos.

Cabe mencionar emplear la investigación experimental, requiere tomar como base los trabajos creativos que son emprendidos de forma sistemática para así aumentar el conjunto de conocimientos con el fin de concebir nuevas aplicaciones, es decir que, el investigador busca comprender fenómenos y procesos, más que acumular datos, lo que concuerda con Muro (2017) y además lo hace con el *método dialéctico* de investigación-reflexión-acción, en donde parte de la experiencia con que la somete a cuestionamiento y la reelabora. Lo que conduce a afirmar, que la función principal de la investigación es sensibilizar y lograr que todos sean conscientes de los problemas.

Por otro lado, se tiene perfectamente delimitado que para innovar hay que aprender y para aprender hay que absorber, por lo que la base del emprendimiento se centra en el hecho de proporcionar al estudiante del Centro Universitario de los Altos (CUAltos), la mejor preparación académica que se tenga en ese momento, con ellos será más competitivo en el momento de su egreso. Ello está basado en el hecho de que en la actualidad, una gran cantidad de instituciones educativas de nivel superior que pertenecen a la Asociación Nacional de Universidades e Institutos de Educación Superior (ANUIES) y cuentan con una plantilla docente que se destaca por la preparación académica obtenida, misma que les permite ser parte del Sistema Nacional de Investigadores (SNI), que como lo establece el CONACYT (s.f.), lo que le obliga al investigador-docente formar parte de una plantilla en un Centro de Investigación o una institución educativa, además de contar con investigaciones publicadas en revistas de alto impacto.

Las instituciones en donde laboran esos investigadores adquieren gran prestigio debido a que durante su estancia, imparten clases a nivel profesional y postgrado, realizan investigaciones en donde involucran a estudiantes (que les permiten adquirir habilidades para la investigación) y realizan publicaciones en revistas de alto impacto, pero sobre todo, que cuentan con invenciones que son patentables.

Por su parte, el CUAltos que forma parte de la red de Centros Universitarios de la Universidad de Guadalajara, cuenta con el Centro de Investigación e Innovación para las Organizaciones (CIIO), que realiza investigaciones orientadas a la obtención de nuevos productos y servicios (y mejorar de los existentes), innova en las metodologías estratégicas de marketing, propicia el emprendimiento, asesora empresas, empresarios de la región (principalmente), estudiantes para que tomen parte de concursos, exposiciones, entre otras tantas actividades. Por tal motivo, el personal que está al frente, son investigadores que cuentan con la preparación requerida para prestar tales servicios.

Todo está relacionado con la nueva estrategia el CIIO, misma que incluye un nuevo indicador como lo es propiciar que el estudiante desarrolle su capacidad para emprender, y alentar al emprendedor con la debida asesoría, para que pueda descubrir si su emprendimiento puede ser conducido a la obtención de la correspondiente patente y comercialización usando técnicas de marketing para que el producto o servicio, sea considerada como exitosa.

Pero tal vez, tenga mayor impacto entre el estudiantado, el hecho de que se esté impulsado al estudiante para que sea capaz de obtener alguna patente o colaborar activamente como parte del equipo que universitario que dirigido por uno de sus investigadores la obtiene. Todo ello está fundamentado se debe a que entre mayor sea su número, será más fácil obtener recursos económicos para los siguientes emprendimientos. Lo que concuerda con lo descrito por autores como Boldrin & Levien (2013), ya que hacen mención al ser mayor el número de patentes obtiene una universidad, mayor es la oportunidad de obtener más recursos y captar el interés de empresas internacionales que puedan invertir en el país.

El motivo para que se obtenga la patente, como parte de los beneficios de la ciencia se debe a que con cada invención patentada, se generan más conocimientos, mismos que los adquiere el estudiante en formación de investigador que esté participando en dicho proyecto, que es un concepto expresado por Bueno y Casani (2017), también se verá beneficiada la compañía tecnológica que lo adquiere e incluya para realizar su trabajo, de tal manera que sea a menor costo, más eficiente y con menos desperdicio.

Por ello, la importancia de la investigación conjunta con el marketing, pueden convertirse en la piedra filosofal del emprendimiento en las universidades, ya que su inclusión de forma ordenada y sistemática, son la base para que el estudiante se convierta en un investigador en potencia al participar en proyectos emprendimiento y bajo la asesoría de investigadores, y las invenciones puedan ser analizadas para potencializarlas y ser sujetas a obtener su patente y ser comercializado de forma exitosa a través del marketing.

Bajo esta premisa, la investigación al impulsar el emprendimiento de proyectos centrados en el estudiante, adquiere mayor relevancia en las presentes y subsecuentes generaciones de profesionistas, ya que a su egreso, llegarán a la empresa, y serán responsables de optimizar sus procesos haciendo uso de las nuevas tecnologías.

Para lograr dicho objetivo, de acuerdo con autores como Loi y Di Guardo (2015), mientras un estudiante forme parte del sistema universitario, su formación no debe de restringirse al aula y laboratorios, y éstos, deben de estar lo más equipado posibles, ya que se busca que se les dote de la mejor preparación académica posible. Así mismo el estudiante, debe ser estimulado para que desarrolle las habilidades que le posibiliten analizar y proponer mejoras a los sistemas conocidos, también, debe adquirir la habilidad de convertir las ideas en proyectos y éstos en emprendimientos, mismos que no se incluyen en las retículas actuales.

Por ello, el presente proyecto está basado en la inducción del estudiante a la investigación, pero a diferencia de cualquier otro método, se hace a través de las patentes que obtiene el investigador-docente y potencializar el mercado mediante el uso de las técnicas marketing más recientes. Cabe hacer hincapié que la preparación actual del egresado del CUAltos es la apropiada, pero se pretende que sea mejor, incorporando conocimientos sobre emprendimiento, patente, marketing y tecnologías de punta, además de los que están incluidos en su retícula.

Por lo que el presente proyecto, tiene su origen en el hecho de que las IES, de acuerdo con los autores Pertuz, Miranda y Sánchez (2021), es el espacio más apropiado para dotar al estudiante con los conocimientos relacionados no solo con las tecnologías de punta, sino por el contrario, con la capacidad de análisis crítico para evaluar, proponer y modificar lo que se esté empleando en la actualidad en la empresa, es decir, que a través participar con docentes-investigadores, sean capaces de ser emprendedores, para que adquieran los conocimientos a partir del emprendimiento, hasta llevarlo a todos los escenarios obligados para la obtención de su patente y comercialización. Todo ese proceso, involucra la adquisición de experiencia que le solicitarán en la empresa, y que, mediante ella podrá asegurar su crecimiento personal y permanencia en la empresa.

Para satisfacer la creciente demanda por mejorar los procesos educativos centrados en el estudiante, el CIIO instrumenta diversos programas de actividades que como programa piloto, estuvo orientado a ser el un plan maestro semestral (incluido el periodo de verano), y con los resultados que se tuvieran, hacer los ajustes y proyecciones necesarias para que se fuera enriqueciendo e incrementando su alcance y cobertura, buscando el beneficio del estudiante (de nivel licenciatura y preparatoria), docente, investigador, empresario, empresa y público en general de la región de los Altos de Jalisco que es donde se encuentra enclavado el CUAltos, sin olvidar que debe expandirse a otras regiones del propio estado y sus vecinos. Así también se tiene que el CIIO, es responsable de realizar y promover eventos que permitan la participación de estudiantes, docentes e investigadores del propio centro educativo, y empresarios de la región de Los Altos, que redunden en la transformación de empresas, servicios y productos, así como en las nuevas empresas que a través de asesorías y participación directa de los diversos integrantes del CUAltos logren eficientar sus procesos de trabajo, productos y hasta el desempeño de su personal.

Por tal motivo, el CIIO trabaja a diario para estar posicionado como un centro de investigación que sea pieza clave en el desarrollo de la región de los Altos, por lo que investigadores, docentes y egresados, están en franca colaboración con los empresarios de la región, para que la calidad educativa que se imparte en él, beneficie directamente a todos aquellos que se preparen académicamente en él, y empresas y empresarios resulten ganadores al recibir a todos aquellos que les ayuden a crecer en forma armoniosa con el objetivo de acelerar el crecimiento de la región.

Para que el CIIO pueda instrumentar las actividades orientadas a cambiar la forma en que piensa un estudiante de acuerdo como lo exponen autores como Vidal y Fernández (2015), se requiere que generar conocimientos centrados en las necesidades propias del estudiante, mismo que tal vez nunca pensó que podría llegar a convertirse en emprendedor e inventar un dispositivo, un método de trabajo o mejorar un dispositivo o método ya existente para hacerlo único y diferente al resto de los de su clase.

Lo anterior, se relaciona con el hecho de que todos los esfuerzos del CIIO, están orientados a impulsar a docentes, estudiantes e investigadores, lo que está justificado por ser de vital importancia para la innovación tecnológica y así contribuir al desarrollo del país. Sin embargo, se esta consiente de que en este momento, de las IES ninguna, fue dotada con la infraestructura y el presupuesto necesarios para incrementar significativamente el número de investigaciones que conduzcan a la solicitud de patentes nuevas. Por ello, la Propiedad Intelectual (PI), una de las políticas que ha adoptado el CIIO, es hacer del conocimiento de todos, que la PI, es una herramienta de negocios y patentes, marcas y otras invenciones considerados como indicadores reales de competitividad empresarial en los sectores productivo y de servicios, así como en las IES.

Por lo que se debe hacer hincapié en el hecho de que, a medida que las IES protejan una mayor cantidad de innovaciones, ideas y creaciones, serán identificadas como instituciones pioneras, competitivas en innovación y transferencia tecnológica, motivo por el cual se debe promover la cultura de la PI en el aula, y romper los paradigmas en la generación de ciencia y tecnología aplicada en la empresa.

Sin embargo, en el CIIO, actualmente existen servicios otorgados a la comunidad universitaria, como son la asesoría y acompañamiento a estudiantes, académicos e investigadores en materia de trámites de solicitudes de invenciones y signos distintivos, protección de derechos, registro de marcas institucionales, además de que se les hace saber la importancia que tiene la PI y sus beneficios en generación no solo de recursos financieros, sino de las estrategias para su difusión, ya que en gran medida dependerá el incremento que se tenga en el número de patentes que inicien desde simples emprendimientos.

Por ello, es menester que todo investigador, tome en consideración que como parte de los conocimientos que comparta con el estudiante, está los esfuerzos que se han realizado para impulsar la investigación con el objetivo de incrementar el número de patentes que realizan los mexicanos. Como parte de esas leyes que debe de conocer cada inventor, está la que se expidió el 1 de julio de 2020, misma que se publicó en el DOF (2020) y es la *Ley Federal de Protección a la Propiedad Industrial* (LFPPI), que entró en vigor en noviembre 5 del 2020. En esa misma fecha también se publicó en el DOF (2020) el decreto por el que se reforman y adicionan diversas disposiciones de la *Ley Federal del Derecho de Autor* (LFDA), para armonizar y adecuar la legislación mexicana en materia de PI con respecto a las obligaciones asumidas por México conforme al Tratado entre México, Estados Unidos y Canadá (T-MEC).

De igual manera, se le debe enseñar al estudiante y tener presente el investigador, que el *tiempo de protección* para modelos de utilidad pasó de *10* a *15* años. Se incorporaron los productos artesanales en los diseños industriales, además de que elimina la obligatoriedad de inscribir las licencias de exportación para que surtan efectos ante terceros. Además, el plazo de vigencia de los registros de marcas se *contará a partir de la fecha de su otorgamiento*, y no desde la presentación de la solicitud IMPI (2021).

Por ello, el CU Altos y específicamente el CIIO, de acuerdo con lo que exponen Degl'Innocenti, Matousek y Tzeremes (2019), se centran en el hecho de que al estudiante se le prepara para el futuro, y debe de estar provisto con las mejores herramientas que podrá usar en el trabajo que desempeñará como profesionista, o bien impulsarlo como emprendedor de su propia empresa y tenga la oportunidad de planearla y construirla desde sus cimientos hasta ser competitiva en su región y a cualquier nivel, mediante el marketing a través del uso de las nuevas tecnologías de la información y comunicaciones.

Para lograrlo el CIIO continúa instrumentando estrategias que permiten al estudiante, docente e investigador, involucrarse en un modelo educativo cultural para superar los obstáculos que se presenten y sea posible incluir dentro de esa cultura formativa el emprendimiento que sustentado en las patentes que son potencializadas por el marketing, sean la base del desarrollo que impulse nuevas empresas con crecimiento sostenido.

## Metodología

La presente investigación está orientada a fomentar al docente e investigador para que a través de sus investigaciones (patentables preferentemente) y usando el marketing en el CU Altos, impulsen el emprendimiento para que se pueda formalizar como parte del modelo educativo actual, como se expresa en Díaz (2015). Con tal propósito, en el CIIO se desarrolla la estrategia para fortalecerlo constantemente, y se le incluyen las actividades que propicien y fomenten el emprendimiento participativo en donde participan estudiante, docente e investigador, con el objetivo de que los productos que se obtengan puedan ser patentables de preferencia; otra estrategia basada en asesorías (se incluye la participación del estudiante) a las empresas de la región de los Altos, y así el participante vea los resultados de los esfuerzos que realiza el CIIO mediante sus programas de vinculación tal como ha sido expresado por Saad y Zawdie (2011).

Para establecer sus estrategias, parte de la perspectiva dialéctica cuyos objetivos han sido adoptados por el CIIO y concuerda con lo establecido por Díaz (2019), ya que a través de la participación del investigador y estudiante, se logra implementar los programas de actividades de investigación que facilitan la comprensión de los fenómenos sociales de la región, y, a través del emprendimiento se logra una participación dinámica para brindar soluciones a los problemas empresariales de la región. Esto está sustentado en el hecho de propiciar soluciones a los problemas de crecimiento y desarrollo sostenibles regionales, por lo que se propone al emprendimiento como factor de cambio.

Por tal motivo, se toma como punto de partida la perspectiva dialéctica, que concuerda con uno de los tantos motivos de ser del CIIO y concuerda con lo establecido por Díaz (2019), ya que a través de la participación que tienen sus investigadores y estudiantes, se pueden implementar los programas de actividades de investigación que faciliten la comprensión de los fenómenos sociales de la región y mediante el emprendimiento tener una participación dinámica para brindar soluciones a los problemas empresariales de la región.

Para solucionar problemas de crecimiento y desarrollo sostenibles regionales, se propone al emprendimiento como factor de cambio. Por ello, los programas de actividades que se planeen en el CIIO deben de ir más allá de ser descriptivo del entorno empresarial y su comportamiento, por el contrario, estar encaminados a identificar, imaginar, proponer y emprender las acciones que permitan al estudiante, docente, investigador y empresario, hacer el uso apropiado del conocimiento y tendencias tecnológicas, para con ello pasar de ser simples espectadores a convertirse en transformadores de empresas e impulsores de los habitantes de la región de los Altos.

Por ello, es que como parte de las actividades que se tienen que renovar día a día en el CIIO, está la de establecer la visión que permita articular la práctica y la teoría a través del emprendimiento, para con ello descubrir, innovar o incorporar las nuevas tecnologías, mejorar procesos y métodos de trabajo y emplear los dispositivos que propicien el crecimiento sostenido del capital humano regional y su industria es por ello que se estableció como estrategia de mercado certificar el CIIO bajo la norma ISO 9001-2015 identificando las siguientes ventajas competitivas:

1.- Primer Centro de Emprendimiento de la Red Universitaria en certificarse con ISO.

2.- Alinear los procesos como empresa de clase mundial.

3.- Brindar confianza y seguridad a los clientes.

4.- Valor agregado ante la competencia.

5.- Establecer el estándar de calidad en los procesos del CIIO.

6.- Facilitar el cumplimiento de objetivos y metas.

7.- Mejorar el proceso de ventas y captación de clientes.

8.- Estandarizar y mantener actualizados los procesos internos.

9.- Abonar a la acreditación de todos los programas académicos del centro universitario.

Debido a ello, muchas actividades se renuevan día a día en el CIIO, lo que se basa en la inclusión de tecnología que permite tener una visión más efectiva para articular la práctica y teoría a través del emprendimiento, y propiciar el descubrir, innovar e incorporar las nuevas tecnologías, mejorar procesos y métodos de trabajo, mejorar y emplear los dispositivos que propicien el crecimiento sostenido del capital humano regional y su industria.

## Objetivo

Diseñar un modelo basado en la influencia que tienen patentes y marketing en el emprendimiento del estudiante del CUAltos para generar proyectos multidisciplinarios.

### Objetivo general

Establecer un modelo, en donde se incorporen docente, investigador, estudiante, empresario y público en general, para promover e impulsar la cultura del emprendimiento a partir de la investigación.

### Objetivo específico

Generar un modelo que propicie e impulse al estudiante para desarrollar habilidades y adquirir conocimientos que faciliten el emprendimiento. Fomentar la implementación de por lo menos *tres* emprendimientos, en donde se incorpore un mínimo de 30 estudiantes. Propiciar que estudiantes del CUAltos tomen parte en concursos regionales y nacionales en áreas del marketing y emprendimiento. Asesorar empresas de la región y organizar eventos en donde participen instituciones educativas de diversos niveles.

lograr la certificación de la norma ISO 9001-2015 como estrategia de mercado que de seguridad al mercado de clientes.

## Desarrollo

Para implementar este modelo de enseñanza basado en la influencia que ejercen las investigaciones que logran la patente, a las cuales se les aplican técnicas de marketing para expandir el alcance y penetración del producto, se hace uso de un modelo experimental que es capaz de permitir hacer las modificaciones necesarias para mejorar y aumentar su penetración en el mercado. Podría decirse que este modelo experimental, ha sido empleado por bastante tiempo, pero en las condiciones actuales en que se encuentra la educación en el país, después de un lapso de tiempo en que la educación se vio afectada por las condiciones de pandemia que imperaron mundialmente, surge como una solución que puede ser la más apropiada tratando de recuperar y tal vez superar las condiciones educativas y de investigación que se tenían antes de ese fenómeno. Por supuesto que, se le aúna el hecho de que en México, como presenta el INEE (2019), el poco favorable equipamiento de aulas y laboratorios, mismo que juega un papel preponderante de manera conjunta con el número de horas que el estudiante asiste a la escuela ya que es variable y solamente durante cuatro o máximo cinco días por semana, lo cual va en contra del número de horas que demanda una investigación, misma que, generalmente, se realiza en horarios fuera de clase y se incluyen fines de semana y días festivos hasta que se tengan los resultados planeados.

Para enfrentar ese problema, el CIIO analizó los factores que debe enfrentar el estudiante durante su formación académica de nivel profesional, y como resultados se estableció que ninguna carrera de nivel licenciatura en las IES (públicas y privadas) tiene asignaturas de emprendimiento e impulso de la investigación que propicien el desarrollo de invenciones susceptibles a ser patentables. Por lo anterior, el CUAltos ha tenido como una de sus tareas principales como parte del funcionamiento del CIIO, fortalecer la detección de las formas en que el estudiante adquiere el conocimiento, y así usando diversas estrategias, docentes e investigadores los refuercen y dirijan al emprendimiento y por ende, se incremente la obtención de patentes, ya que no todo emprendimiento resulta patentable.

Por lo anterior, el CUAltos ha tenido como una de sus tareas principales como parte del funcionamiento del CIIO, fortalecer en primera instancia detectar la forma en que los conocimientos que adquieren sus estudiantes, docentes e investigadores pueden ser reforzados y dirigidos hacia implementación de programas que permitan realizar emprendimientos y por ende, la obtención de patentes.

Por tal motivo, uno de los programas en lo que más hincapié se ha hecho en el CIIO, es el que se ha instrumentado y enfocado a propiciar e incrementar el emprendimiento entre estudiantes, docentes e investigadores, ya que son ellos en quienes recae la responsabilidad de fortalecer la industria de la región de Los Altos, lo que está basado en el desarrollo o mejoramiento de métodos, técnicas y dispositivos novedosos e innovadores, que puedan ser aplicados en las empresas sin que se pierda el objetivo de que puedan llegar a adquirir la calidad de ser sometidas a la revisión del IMPI (Instituto Mexicano de la Propiedad

Industrial) para que le sea concedida la patente y solamente pueda ser usada por quien se haga acreedor a la misma.

Otro de los programas que ha instrumentado el CIIO que además de estar enfocado en el emprendimiento entre sus estudiantes, es la implementación del método *para la nueva Era* que ha sido explicado por Velez (2016) como una creación de Steve Jobs en el se propone un aprendizaje autónomo, modelo que se tuvo la necesidad emplear durante la época de la pandemia de SARS-COV2, ya que durante ese tiempo, el estudiante del CUAltos, se vio obligado a establecer sus propias metas guiado por el docente y apoyados por el personal y programas del CIIO.

Como resultado, se ha hecho hincapié en que el CIIO se convierta en un instrumento enfocado en propiciar e incrementar el emprendimiento entre estudiantes, docentes e investigadores, para que sea el modelo que fortalezca la industria de la región de Los Altos, basado en el desarrollo o mejoramiento de métodos, técnicas y dispositivos novedosos e innovadores, que puedan ser aplicados en las empresas. Así también, los productos que se obtengan, sean sometidos a la revisión del Instituto Mexicano de la Propiedad Industrial, para que le sea concedida la patente y solamente pueda ser usada por quien se haga acreedor a la misma. Sin olvidar que un papel preponderante por el grado de influencia que tiene, es el marketing, ya que sino se tienen las estrategias apropiadas para darlo conocer y hacerlo llegar al consumidor, entonces correrá el riesgo ser olvidado rápidamente.

Debido a al premura con que se autorizó el regreso a actividades presenciales, en el próximo pasado mes de enero, tal como lo reportó Montiel (2022), se tuvo que implementar aceleradamente un programa general de actividades para el primer periodo del presente año lectivo, en él se incluyó el verano con el objetivo de cubrir algunas actividades que se podrían realizar en ese periodo. Por lo que las estrategias se que se instrumentaron, se clasificaron en seis rubros, los cuales son:

*1 Estrategias de expansión educativa*: Organizar y participar en congresos (workshops), conferencias, talleres, concursos de emprendimiento, diplomados, reuniones de trabajo con instituciones educativas y de gobierno para expandir los servicios que presta el CIIO.

*2 Estrategias de vinculación*: Organizar y participar en encuentros entre directivos con estudiantes, investigadores, entidades corporativas, autoridades municipales y estatales.

*3 Estrategias para fomentar el emprendimiento*: Asesorar proyectos de emprendimiento internos y externos.

*4 Estrategias para consolidar proyectos*: Brindar asesoría jurídica para la constitución empresarial, registro de marcas y patentes para registro ante el IMPI.

*5 Estrategias para incrementar la influencia del CUAltos*: Dar a conocer las actividades que realiza el CIIO a estudiantes de nuevo ingreso, y de otros niveles educativos, empresarios e instituciones que puedan en determinado momento interesarse en ello.

*6 Estrategias de capacitación*: asesorar a estudiantes para participar en concursos de conocimientos en las áreas de emprendimiento y marketing.

*7 Estrategia de certificación bajo la norma ISO 9001-2015 los procesos de:* Procesos de Investigación, Capacitación, Asesoría / Consultoría, Incubación y Promoción de Proyectos.

## Resultados

En la Tabla 1, se muestran las estrategias que se realizaron en el mes de febrero del 2022, las cuales quedaron agrupadas de las siguiente manera.

**Tabla 1.** Estrategias implementadas en el mes de Febrero agrupadas acorde a su orientación.

| Estrategia | Febrero |
|---|---|
| 1 | Un workshop para orientar y proporcionar herramientas a los participantes para que propongan e implementen soluciones innovadoras que transformen la realidad y generen valor social, ambiental y económico, a través de emprendimientos factibles y viables. |
| 3 | Asesoría de marketing para uno de los proyectos de marketing que se emprendieron. |
| 3 | Asesoría a 15 estudiantes para que generen proyectos de emprendimiento. |
| 4 | Asesoría jurídica para llenar contratos laborales del emprendimiento. |
| 5 | Visita de industriales para darles a conocer las actividades del CIIO, |
| 5 | Atención a 90 estudiantes de preparatoria, para darles a conocer las actividades que se desarrollan en CIIO. |
| 5 | Atención a 35 estudiantes de primer ingreso de la carrera de administración, para darles a conocer las actividades que se desarrollan en CIIO. |

Fuente: Elaboración propia con datos del CIIO

En la Tabla 2, se muestran las estrategias que se realizaron en el mes de marzo del 2022, las cuales quedaron agrupadas de las siguiente manera.

**Tabla 2.** Estrategias implementadas en el mes de Marzo agrupadas acorde a su orientación

| Estrategia | Marzo |
|---|---|
| 1 | Se realizó una conferencia sobre emprendimiento para motivar al estudiante en general a que de inicio con su propio emprendimiento. |
| | Se efectuó una reunión con los demás integrantes de la Red de Centros Universitarios de la U de G, para establecer un comparativo de resultados en materia de emprendimiento. |
| 2 | Atender al personal de los municipios en el área de influencia del CUAltos y darles a conocer las actividades que desarrolla el CIIIO e identificar los proyectos que pueden ser realizados en conjunto con ellos o para ellos. |
| 3 | Dio inicio con el análisis de las actividades que se involucran para comprobar el funcionamiento de un emprendimiento que se estuvo realizando durante el 2022. |
| | Se dio asesoría de dos proyectos de emprendimiento. |
| | Atender a 16 emprendedores para establecer las condiciones para la cooperación entre ellos y CU Altos. |
| | Se brindaron 14 sesiones de asesoría a cuatro proyectos de emprendimiento internos y uno externo. |
| | Se dio asesoría para que inicie un nuevo proyecto de emprendimiento. |
| | Se dieron dos sesiones de asesoría a dos equipos de incubadoras de proyecto para iniciar sus actividades. |
| 5 | Se atendió dos grupos (64) estudiantes de preparatoria para explicarles las actividades que realiza el CIIO. |
| 6 | Se brindaron siete sesiones de asesoría durante el transcurso del mes de marzo, a los estudiantes de los equipos que participarán en el concurso de ANFECA en su fase regional, y posteriormente a aquellos que lleguen a la fase nacional. |

Fuente: Elaboración propia con datos del CIIO

En la Tabla 3, se muestran las estrategias que se realizaron en los meses de abril y mayo del 2022, se toman de manera conjunta debido a que en abril, solo se tienen dos semanas de actividades por el periodo vacacional de primavera y mayo completo, y quedaron agrupadas de las siguiente manera.

**Tabla 3.** Estrategias implementadas en abril y mayo, se agrupan acorde a su orientación

| Estrategia | Abril – Mayo |
|---|---|
| 1 | Estructuración de la organización para ser anfitriones del concurso regional de ANFECA, por lo que hicieron los programas de actividades para llevar a efecto y resultara una buena experiencia para la gente local (personal, estudiantes, personal administrativo, docente y de investigación y empresarios de la región de Los Altos) y los visitantes que tomarían parte en el mismo, como competidores, asesores e interesados en conocer sus proyectos. |
| | Se organizaron tres encuentros de estudiantes emprendedores del CUAltos, una con directivos del propio centro y dos de esos encuentros de vinculación, siendo uno de ellos con el INADEJ y otro con empresarios de la región. |
| 4 | Se dieron tres asesorías sobre registro de marcas y patentes, además de asesorías a dos distintas empresas de la región para la mejora de sus procesos, incluida la capacitación. |
| 5 | Tres eventos de difusión del concurso de ANFECA invitando a todo el personal y estudiantes del CU Altos, así como gente (empresarios y público en general) que puedan interesarte por iniciar a involucrarse en el mundo de los negocios, hacia donde esta orientado el concurso. |
| | También se aprovecharon los eventos en que participan estudiantes del CUAltos para dar a conocer las actividades que realiza el CIIO y tratar de que más gente pueda solicitar los servicios que brinda. |
| | También se organizó y participó en un encuentro entre CUAltos/CIIO/COPARMEX/Autoridades municipales. |
| 6 | Se tuvo un total de 10 asesorías relacionadas con el concurso de ANFECA, asesorando a los equipos participantes por parte del CUAltos, del comité técnico, así como la celebración del propio concurso. |
| | Se brindaron dos sesiones de asesoría a los estudiantes que participan en el concurso organizado por el CUCEA, así también se como en el concurso municipal de emprendimiento. |

Fuente: Elaboración propia con datos del CIIO

En la Tabla 4, se muestran las estrategias que se realizaron por el CIIO en los meses de junio y julio del presente año, estos meses se presentan de manera conjunta ya que el semestre escolar concluye a finales de junio y el inicio del mes de julio, por lo cual la mayoría de ellas, se ubican en junio, y quedaron agrupadas de las siguiente manera.

**Tabla 4.** Estrategias implementadas en junio y julio, se agrupan acorde a su orientación.

| Estrategia | Junio - Julio |
|---|---|
| 1 | se tuvieron dos sesiones en donde se atendieron a los participantes de la segunda generación de participantes del diplomado DIDENCE (Desarrollo de Negocios y Cultura del Emprendimiento), ya que a la segunda se dio inicio formal al programa del mismo. |
| | Por último se hizo la organización y presentación de la convocatoria y mecanismos para la participación general del Centro Universitario para obtener apoyos para proyectos, concluyendo el periodo semestral con la presentación de la Agenda de actividades calendario 2022-B para el CIIO. |
| 2 | Se llevó a afecto una sesión informativa y de trabajo sobre los nuevos prestadores de servicio social y prácticas profesionales de estudiantes del CU Altos. |
| 3 | se tuvieron cuatro sesiones de asesoría y evaluación de cuatro proyectos diferentes de emprendimiento, |
| 5 | se realizaron cinco sesiones se evaluación y seguimiento de la norma ISO 9001:2015 |
| 6 | Se tuvo una sesión de asesoría con aquellos estudiantes que participarán el concurso de ANFECA en su fase nacional, |
| | Se realizaron los trámites sobre capacitación para certificación como Agentes para la impartición de cursos en el trabajo alineado a la Secretaría del Trabajo y Previsión Social. |

Fuente: Elaboración propia con datos del CIIO

En la Tabla 5, se muestran las estrategias que se realizaron por el CIIO en los meses de agosto y septiembre del presente año, estos meses se presentan de manera conjunta ya que el semestre escolar a concluido y abarca el verano escolar, y quedaron agrupadas de las siguiente manera.

**Tabla 5.** Estrategias implementadas en agosto y septiembre, se agrupan acorde a su orientación.

| Estrategia | Agosto - Septiembre |
|---|---|
| 1 | Un coloquio con estudiantes, docentes e investigadores de la carrera de Derecho para brindarles la asesoría sobre las actividades, leyes y reglamentos correspondientes a los procesos de registro de marcas. |
| | Un curso de verano infantil orientado a fomentar la participación de los niños de la región para que vayan identificando las actividades que se realizan en el CU Altos. |
| 2 | Un encuentro con COPARMEX Guadalajara para ofrecer los servicios que presta el CIIO. |
| | Se organizó y participó en la reunión que facilitará el establecimiento de un proceso de manufactura por parte del CIIO dirigido a empresas de la región. |
| 4 | Seguimiento de tres de los proyectos de emprendimiento que se han estado asesorando. |
| | Una sesión con los asesores de los proyectos que han de participar en el concurso nacional de ANFECA. |
| | Se organizó y participó en la reunión con el Colegio de Contadores de la Región de los Altos para exportación y procesos contables internacionales. |
| 5 | Tres sesiones a lo largo del mes de agosto para evaluar y hacer entrega de la documentación que se requiere para el proceso de acreditación del ISO 9001:2015. |
| | Se elaboró y presentó el calendario de actividades relacionadas con todas las fases del emprendimiento y que se han de efectuar en el presente periodo escolar universitario. |
| | Continuación del diplomado de DIDENCE. |
| 6 | Cuatro sesiones de asesoría |
| | Una sesión se asesoría para impulsar los emprendimientos que se realizan en la región. |

Fuente: Elaboración propia con datos del CIIO

# Distribución de Resultados

Durante *Febrero*, las estrategias implementadas, se contó con la participación de investigadores del CIIO, así como docentes, empresarios y estudiantes en general, son presentados en la Figura 1,

**Figura 1.** Actividades del CIIO para fomentar el emprendimiento en Febrero de 2022

Fuente: Elaboración propia con las estadísticas del CIIO

De las estrategias, el *43 porciento*, estuvieron orientadas a incrementar la influencia del CU Altos, el *29 porciento* a fomentar el emprendimiento, el *14 porciento* a la expansión educativa y 14 *porciento* a la consolidación de los proyectos.

Durante *marzo*, se tuvieron actividades encaminadas a mostrar que tan competitivo es el CIIO y los resultados de las actividades realizadas en este mes se pueden apreciar en la Figura 2.

**Figura 2.** Actividades Realizadas por el CIIO para fomentar el emprendimiento y registro de patentes de invenciones y marcas en Marzo de 2022

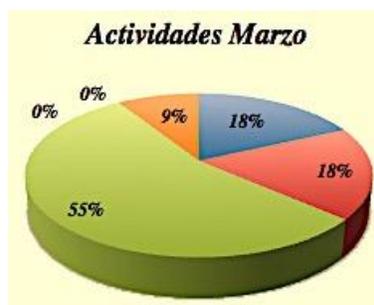

Fuente: Elaboración propia con las estadísticas del CIIO

Del total de las estrategias que se implementaron, el *55 porciento* fue al fomento del emprendimiento. El *18 porciento* a la expansión educativa. El *18 porciento*, a la vinculación y tan solo el nueve porciento a la capacitación.

En abril tan solo se tuvo la oportunidad de laborar dos semanas y mayo, por su parte permitió que tuvieran implementaran a lo largo del mismo, los resultados que se obtuvieron se muestran en la Figura 3.

**Figura 3.** Actividades Realizadas por el CIIO para fomentar el emprendimiento y registro de patentes de invenciones y marcas en Abril–Mayo de 2022

Fuente: Elaboración propia con las estadísticas del CIIO

Del total de las estrategias, el *38 porciento* de ellas, estuvo orientado a incrementar la influencia del CUAltos. El *25 porciento* estuvo relacionado con la expansión educativa, el 25 porciento a la capacitación, por último el 12 porciento a la consolidación de proyectos.

El periodo *junio – julio,* muestra los resultados que se obtuvieron en el periodo y se pueden ver en la Figura 4.

**Figura 4.** Actividades Realizadas por el CIIO para fomentar el emprendimiento y registro de patentes de invenciones y marcas en junio-julio de 2022

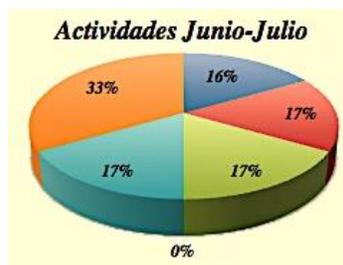

Fuente: Elaboración propia con las estadísticas del CIIO

El 33 *porciento* fueron actividades de las estrategias de capacitación. El 17 *porciento* corresponden a fomentar el emprendimiento, otro 17 *porciento* fueron de las estrategias de expansión educativa y finalmente, el 16 *porciento* fueron para incrementar la influencia del CUAltos.

El último periodo que comprende esta investigación, corresponde al mes de *agosto* completo y la primer parte del mes de *septiembre* (como fueron agrupados los resultados que se pueden apreciar en la Figura 5), lo que reside en la programación de actividades del calendario 2022A de la U de G y CU Altos, mismo que forma parte de la Red de Centros Universitarios de la misma, y por consecuencia, se debe de realizar sus actividades tomando en consideración ese calendario.

Al ser un periodo que se ubica entre la terminación del primer semestre del año y el segundo, se incluyó la organización del verano infantil, en el cual participan los hijos del personal de la institución y de la población en general, ya que se ha desarrollado la política de que entre más pequeños sean, se les debe dar a conocer qué hacen los investigadores, y docentes en lo relacionado con el emprendimiento, para de esta forma despertar su interés por actividades que le ayudarán en el futuro a beneficiar a la región de Los Altos.

**Figura 5.** Actividades Realizadas por el CIIO para fomentar el emprendimiento y registro de patentes de invenciones y marcas en Agosto-Septiembre de 2022

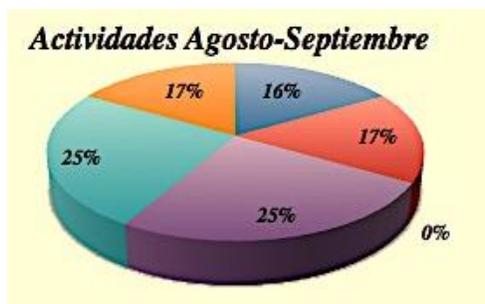

Fuente: Elaboración propia con las estadísticas del CIIO

El 25 *porciento* de las actividades de la estrategia de expansión educativa. Otro 25 *porciento*, fueron encaminadas a incrementar la influencia del CUAltos, mientras que el 17 *porciento* de correspondieron a estrategias de vinculación, otro 17 *porciento* fueron de la estrategias de expansión educativa y solamente el 16 *porciento*, fueron encaminadas a la capacitación, lo cual se puede entender debido a que es el momento en que entra en receso la Universidad.

# Discusión

Se tuvieron que implementar lo más rápido posible para que el investigador del CUAltos, conjuntamente con docentes y personal del CIIO, buscaran incorporar dentro de sus actividades diarias, las correspondientes a mostrar al estudiante y empresario, que una patente, es de suma importancia no solo para quien la obtiene, sino que por el contrario, se benefician todos de la nueva tecnología que se genera dentro del CUAltos, y cuyo objetivo es impulsar a través del emprendimiento, la filosofía de los beneficios de expandir la producción que se hace en la región de los Altos, lo que se ha incrementado desde el momento en que se le han aplicado técnicas de marketing que trabajan sobre las tecnologías de la información y comunicación, para darlas a conocer más allá del alcance que tienen sus distribuidores tradicionales.

Además, se ha tomado en consideración de que se recurrió a la instrumentación de estrategias basándose en las patentes y marketing, con lo que se busco que fuera más amplio a la vez que permitiera extender el área de influencia del CUAltos, ya que la mayoría de sus estudiantes conocen la institución por ser provenientes de la misma región, por lo que a ellos, se les impulsa a que participen en emprendimientos debido a que esos emprendimientos pueden beneficiar a su región, universidad y a ellos mismos.

Además, de que este primer ciclo permite hacer una evaluación del comportamiento de la aplicación de las estrategias que se implementaron, ya que son el primer paso para seguir creciendo en beneficio de la misma institución, estudiantado, personal administrativo y académico, para que finalmente todos esos beneficios se vean reflejados en el crecimiento de la región, misma que se caracteriza por contar con empresas cuyos productos deben de darse a conocer fuera de su área natural de influencia, con ello, se podrá internacionalizar la producción regional, aprovechando las ventajas de participar en eventos nacionales y llegar más allá de las fronteras del país, al estar incursionando en el área de productos patentables que pueden ser comercializado bajo las estrictas reglas del marketing.

Por último, se debe mencionar que Aboites y Díaz (2018), sobre todo esta última, se ha dedicado a realizar investigaciones sobre el comportamiento que ha tenido la concentración de solicitudes y otorgamiento de patentes, haciendo resaltar el hecho de que la Ciudad de México, está es primer lugar, seguida de Jalisco, lo que resulta una forma de suma importancia en el papel que está desempeñando del CUAltos, ya que al formar parte del sistema educativo de la U de G, es de vital importancia que se siga contribuyendo como centro universitario a incrementar el número que se tiene de patentes, con el objetivo de hacer crecer las estadísticas que permitan paulatinamente disminuir la brecha que separa el número de patentes que se tienen entre la Ciudad de México y el Estado de Jalisco.

# Conclusiones

Este primer periodo semestral ha sido una primer prueba que permitió implementar este plan inicial compuesto por estrategias, mismas que se sustentaron en los beneficios en la formación académica que se pueden obtener de una patente que obtenga alguno de los investigadores del CUAltos, ya que es tan solo un aliciente para que el resto de los investigadores empleen la misma técnica en donde puedan incorporar a una mayor cantidad de estudiantes, empresarios y docentes, todos aquellos que busquen propiciar el crecimiento de la región de los Altos.

De igual manera, el hecho de haber incorporado estudiantes para que adquieran experiencia en la asesoría que se proporciona al empresario de la región, es altamente gratificante, ya que por metodología, se les ha adentrado en la aplicación del marketing, desde sus versiones iniciales, para que adquieran la capacidad de emprender sus propios proyectos y proyectarlos de tal manera que, puedan optar con por continuar con su preparación en la investigación, buscando contribuir con la región de donde provienen.

Para el personal del CIIO, la etapa de pandemia, permitió preparar las estrategias que podrían emplearse para fomentar e impulsar las actividades que incrementaran el número de proyectos que, a través del emprendimiento, pudieran contribuir a incrementar el número de patentes que tiene la U de G, la cual se redujo sustancialmente en ese periodo. A través de las estrategias que estuvieron enfocadas a motivar, fomentar y propiciar, mediante las asesorías en diversas áreas del conocimiento, el crecimiento del estudiante en conocimientos adquiriendo herramientas que le faciliten su inclusión en las empresas al ser profesionistas, además de que a través del emprendimiento adquieran la cultura por la mejora de los procesos empresariales.

Sin embargo, se está consiente de que se pudieren haber obtenido más logros de los que se obtuvieron en el presente ciclo escolar, y que son la base para la planeación de las estrategias del próximo. Lo anterior se fundamenta en el hecho de que se tuvo muy poco tiempo para determinar cuáles serían las estrategias que se emplearían para que se integraran los conceptos que se desprenden de las patentes y el marketing, ya que su influencia es de vital relevancia en el crecimiento de la gente que participa en ese tipo de proyectos, así como de la comunidad en general en donde se encuentra enclavado el CUAltos.

Este proyecto fue experimento, que ha permitido darse cuenta de que no se obtuvieron los éxitos que fueran en cantidad deseado, pero si es un incentivo par que se instrumente un plan de trabajo mucho mejor y más ambicioso en sus alcances, ya que todos los participantes, han tenido una excelente sensación de lo que aprendieron, sus expectativas se vieron superadas, por lo que estarán buscando incorporar a otras personas que por algún motivo en este periodo no estuvieron formando parte del proyecto, por lo que el entusiasmo que tiene todo aquel que si participó, es contagioso que se esperará una mayor respuesta por parte de la comunidad en general.

De esta forma se prevé que se aumentará el número de emprendimientos por incubar y realizar más trámites para conseguir patentes, con lo que se podrá continuar influyendo en las nuevas generaciones para que se conviertan en emprendedores que beneficien la región, den mayor prestigio al CUAltos. Por ello, se tiene una satisfacción momentánea al haber cumplido las metas, pero se tiene el acicate de que la meta debe crecer para que haya más y nuevos emprendedores, docentes, investigadores y empresarios que paulatinamente se incorporen a este proyecto.

## Trabajo futuro

Terminar el periodo semestral en donde se logró tener una participación más o menos adecuada en el proyecto, se tiene que replantear completamente el conjunto de estrategias para hacerlas más atractivas y con metas y objetivos aún mayores. Por otro lado, se notó que la participación de investigadores con patentes fue un factor detonante para que se interesaran estudiantes y empresarios, se ha planeado que el marketing de las nuevas generaciones sea aún más incisivo debido a que, se sabe que conseguir una patente no es sencillo, por lo que a la par se deberán de desarrollar las estrategias que permitan proyectar en el tiempo la participación que se irá teniendo paulatinamente por parte de todos los integrantes de cada emprendimiento, ya que es la base para proyectar al CUAltos y ponerlo en los primeros lugares entre los centros universitarios de la red de U de G, lo que es menester para el crecimiento de la región de los Altos.

Además, se debe ampliar las relaciones con los empresarios y emprendedores de la región para establecer convenios y proporcionarles la asesoría que necesiten para sus proyectos y trámites que se realizan para el registro de invenciones, modelos de utilidad, marcas y cualquier otro que esté relacionado con el IMPI.

Sin olvidar que los concursos en que participen los estudiantes del CUAltos, es otro punto medular para influir en la captación de participantes que sean capaces de proponer y desarrollar sus propios emprendimientos. Así se estará fomentando e impulsando el crecimiento de estudiantes emprendedores, empresas que requieran los diversos servicios del CUAltos. Por último, hacer estudios y trámites necesarios para integrar en la retícula de las carreras impartidas en el CUAltos, una materia de emprendimiento.

## Referencias

Aboites, J., y Díaz, C. (2018). Auge y declinación en la producción de conocimiento codificado en patentes en el IMP. Innovación, núm. 1.

Alvarado Reséndiz, J. L., García Munguía, M. y Castellanos López, L.Y. (2017). Aprendizaje Significativo en la Docencia de la Educación Superior. XIKUA Boletín Científico de la Escuela Superior de Tlahuelilpan. Vol. 5 Núm. 9 DOI: https://doi.org/10.29057/xikua.v5i9.2239

Boldrin, M. & Levine, D.K. (2013). The Case against Patents. The Journal of Economic Perspectives. A Journal of the American Economic Association. Volume 27. Number 1. Winter. Ed. David H. Autor, M. I. T. P. 3-22.

Bueno, E., y Casani, F. (2007). La tercera misión de la universidad. Enfoques e indicadores básicos para su evaluación. Economía Industrial, núm. 366, 43-59.

Catalán Cueto, J. P. (diciembre 20, 2020). LA INVESTIGACIÓN ACCIÓN COMO ESTRATEGIA DE REVISIÓN DE LA PRÁCTICA PEDAGÓGICA EN LA FORMACIÓN INICIAL DE PROFESORES DE EDUCACIÓN BÁSICA. Revista Ibero-Americana de Estudos em Educação, vol. 15, núm. 4, Esp., pp. 2768-2776. Universidade Estadual Paulista Júlio de Mesquita Filho, Faculdade de Ciências e Letras

Chesbrough, H., Vanhaverbeke, W. y West, J. (septiembre 30, 2014). Explicando la Innovación Abierta: Aclarando un Paradigma Emergente para Comprender la Innovación. Nuevas Fronteras en Innovación Abierta. Oxford University Press. Oxford.

CONACYT (s.f.). Sistema Nacional de Investigadores. Consejo Nacional de Ciencia y Tecnología. https://conacyt.mx/sistema-nacional-de-investigadores/

Degl'Innocenti, M., Matousek, R., y Tzeremes, N. G. (2019). The interconnections of academic research and universities "third mission": Evidence from the uk. Research Policy, 48(9), 103793.

Díaz, C. (2015). Flexibilidad y autonomía en la generación de conocimiento: la experiencia de la UAM-Iztapalapa. En Aboites, J., y Díaz, C. Díaz. (2015). Inventores y patentes académicas: la experiencia de la Universidad Autónoma Metropolitana, 154-271. Siglo XXI Editores/Universidad Autónoma Metropolitana

Díaz, C. (2019). Ciencia, tecnología e innovación: retos estratégicos de Jalisco. A. Acosta (ed.). Jalisco a futuro 2018-2030. Construyendo el porvenir, Vol. I, 282-336. Ediciones Universidad de Guadalajara.

Diego Bautista, O. (diciembre 2019). Generación y divulgación del conocimiento. El rol del investigador y del repositorio institucional. Revista de Identidad Universitaria, v. 1, n. 7, p. 26-28, ISSN 2448-7651. https://revistaidentidad.uaemex.mx/article/view/13602

DOF. (julio 1, 2020). Diario Oficial de la Federación. Secretaría de Economía. https://www.dof.gob.mx/index_113.php?year=2020&month=07&day=01#gsc.tab=0

DOF. (julio 1, 2020). Diario Oficial de la Federación. Secretaría de Cultura. https://www.dof.gob.mx/index_113.php?year=2020&month=07&day=01#gsc.tab=0

IMPI. (2021). impi en Cifras 2020: 1993 a septiembre 2020. https://www.gob.mx/impi/documentos/instituto-mexicano-de-la-propiedad-industrial-en-cifras-impi-en-cifras.

INEE (2019). Panorama educativo de México: Indicadores del Sistema Educativo Nacional. México: INEE.

Loi, M., y Di Guardo, M. C. (2015). The Third Mission of Universities: An Investigation of the Espoused Values. Science and Public Policy, 42(6), 855-870.

Mackay Castro, R., Franco Cortazar, D. E. y Villacis Pérez, P.W. (enero-marzo 2018). El pensamiento crítico aplicado a la investigación. Universidad y Sociedad vol.10. no.1. Cienfuegos. Quito, Ecuador. ISSN 2218-3620

Montiel, A. (enero 11, 2022). La Universidad ajusta su fecha de regreso a clases presenciales: Para Prepas UDG y primer ingreso de licenciaturas será el 31 de enero. U de G. https://www.sems.udg.mx/noticias/la-universidad-ajusta-su-fecha-de-regreso-clases-presenciales-para-prepas-udg-y-primer

Muro, R. del. (septiembre 2017). Aportes del método materialista dialéctico para fortalecer la intervención profesional con reflexiones sobre la metodología situada en lo concreto. SEDICI. Repositorio Institucional de la Universidad Nacional de La Plata. http://sedici.unlp.edu.ar/handle/10915/64333

OCDE (2015), Frascati Manual 2015: Guidelines for Collecting and Reporting Data on Research and Experimental Development, The Measurement of Scientific, Technological and Innovation Activities. Publicado por acuerdo con la OCDE, París (Francia). DOI: http://dx.doi.org/10.1787/9789264239012-en

Pertuz, V., Miranda, L.F. y Sánchez Buitrago, J. O. (octubre 12, 2021). Innovación tecnológica en educación: una revisión de literatura sobre los MOOC desde la perspectiva docente. REVISTA INTERAMERICANA DE INVESTIGACIÓN, EDUCACIÓN Y PEDAGOGÍA. https://revistas.usantotomas.edu.co/index.php/riiep/article/view/7856/7489

Poblete, F., Linzmayer, L., Matus, C., Garrido, A., Flories, C., García, M., & Molina, V. (2019). Enseñanza-Aprendizaje basado en investigación. Experiencia piloto en un diplomado de motricidad infantil. Retos (35), 378-380.

Rivadeneira, E. y Silva, R. (2017). Aprendizaje Basado en Investigación en el trabajo autónomo y en equipo. Negotium, 13(38), 5-16.

Ruiz de Olano, D. & Ageitos, N. (septiembre 2013). Para innovar con éxito es necesario aprender. Business Review. Habilidades directivas. https://www.harvard-deusto.com/para-innovar-con-exito-es-necesario-aprender

Saad, M. y Zawdie, G. (2011). Introduction to Special Issue: The Emerging Role of Universities in Socio-Economic Development through Knowledge Networking. Science and Public Policy, 38(1), 3-6.

Sciarelli, M., Gheith, M. H. y Tani, M. (2020). The relationship between quality management practices, organizational innovation, and technical innovation in higher education. Quality Assurance in Education, 28(3), 137-150. https://doi.org/10.1108/QAE-10-2019-0102

Tecnológico de Monterrey. (2020). Aprendizaje Basado en la Investigación. Innovación Educativa en el Tecnológico de Monterrey. http://innovacioneducativa.tec.mx/aprendizaje-basado-en-investigacion/

Terrazas Pastor, R. y Silva Murillo, R. (2013). La educación y la sociedad del conocimiento. Perspectivas, Año 16 – Nº 32 – octubre 2013. pp. 145-168. Universidad Católica Boliviana "San Pablo", Unidad Académica Regional Cochabamba. https://www.redalyc.org/pdf/4259/425941262005.pdf

Tovar Góngora, Á. E. (julio 23, 2014). El impacto de las TIC'S en el marketing. Gestiopolis. https://www.gestiopolis.com/el-impacto-de-las-tics-en-el-marketing/

Vázquez Parra, J.C. (julio 2, 2021). ¿Cómo detonar el Aprendizaje Basado en Investigación en el Aula?. Institute for Future Education. Tecnológico de Monterrey. Observatory. https://observatorio.tec.mx/edu-bits-blog/aprendizaje-basado-en-investigacion/

Vidal L. M.J. y Fernández O., B. (2015). Aprender, desaprender, reaprender. Scielo. SSN 1561-2902. Educ Med Super vol.29 no.2 Ciudad de la Habana abr.-jun. 2015. http://scielo.sld.cu/scielo.php?script=sci_arttext&pid=S0864-21412015000200019